\documentclass[fleqn,usenatbib]{mnras}

\usepackage{newtxtext,newtxmath}

\usepackage[T1]{fontenc}
\usepackage[utf8]{inputenc}

\DeclareRobustCommand{\VAN}[3]{#2}
\let\VANthebibliography\thebibliography
\def\thebibliography{\DeclareRobustCommand{\VAN}[3]{##3}\VANthebibliography}

\usepackage{graphicx}	
\usepackage{amsmath}	

\title[Radio and X-ray observations of Swift J1818.0-1607]{Long term radio and X-ray evolution of the magnetar Swift J1818.0-1607}

\author[K.M. Rajwade et al.]{
K. M. Rajwade,$^{1}$\thanks{E-mail: kaustubh.rajwade@manchester.ac.uk (KMR)}
B. W. Stappers,$^{1}$
A. G. Lyne, $^{1}$
B. Shaw,$^{1}$
M. B. Mickaliger,$^{1}$
K. Liu, $^{2}$
M. Kramer, $^{2}$
\newauthor
G. Desvignes, $^{2}$
R. Karuppusamy $^{2}$
T. Enoto,$^{3}$
T. G\"{u}ver,$^{4,5}$
Chin-Ping Hu$^{6}$
and M. P. Surnis $^{1}$\\
$^{1}$Jodrell Bank Centre for Astrophysics, University of Manchester, Oxford Road, Manchester M13 9PL, UK\\
$^{2}$Max-Planck-Institut f\"ur Radioastronomie, Auf dem H\"ugel 69, D-53121 Bonn, Germany \\
$^{3}$ RIKEN Cluster for Pioneering Research, 2-1 Hirosawa, Wako, Saitama 351-0198, Japan\\
$^{4}$Istanbul University, Science Faculty, Department of Astronomy and Space Sciences, Beyazıt, 34119, Istanbul, Turkey\\
$^{5}$ Istanbul University Observatory Research and Application Center, Istanbul University 34119, Istanbul Turkey\\
$^{6}$Department of Physics, National Changhua University of Education, Changhua 50007, Taiwan\\
}

\date{Accepted XXX. Received YYY; in original form ZZZ}

\pubyear{2015}

\begin{document}
\label{firstpage}
\pagerange{\pageref{firstpage}--\pageref{lastpage}}
\maketitle
\begin{abstract}
We report on the the long term monitoring campaign of the seemingly youngest magnetar
Swift~J1818.0-1607 at radio and X-ray wavelengths over a span of one year. We obtained a
coherent timing solution for the magnetar over the same time span. The frequency
derivative of the magnetar shows systematic variation with the values oscillating about a
mean value of $-$1.37$\times$10$^{-11}$~Hz~s$^{-1}$. The magnitude of the variation in the
frequency derivative reduces with time before converging on the mean value. We were able
to identify four states in the spin-frequency derivative that were quantified by the
amount of modulation about the mean value and the transition between these states seem to
be correlated with the change in the radio emission of the magnetar while no correlation
is seen in the average radio profile variability on a shorter timescale (days). The
0.5--12~keV X-ray flux shows a monotonic decrease that can be attributed to thermal
emission from a hot-spot on the surface of the neutron star that is reducing in size. Such
decrease is consistent with what is seen in other magnetars. The potential correlation
between the radio emission mode and the behaviour of the spin-down rate hints to a global
change in the magnetopshere of the magnetar akin to the correlation seen in a subset of
mode-changing radio pulsars and suggests a physical link between the two sub-populations.

\end{abstract}

\begin{keywords}
stars:neutron -- pulsars:general -- stars:magnetars
\end{keywords}



\section{Introduction}
Neutron stars are one of the most enigmatic astrophysical objects known over the last 50 years ago~\citep{hewish1968}. Magnetars are a sub-class of neutron stars and are characterized by high magnetic fields (B$\geq$10$^{14}$~G), long spin periods (1--12 seconds) and small characteristic ages~\citep[see][and the references therein]{kaspi2017}. They are distinguished from the radio pulsars by their high-energy emission that is thought to be powered by the magnetic field energy rather than the rotational energy of the neutron star that is powering radio pulsars. Furthermore, they exhibit this high-energy emission in the form of episodic short (a few seconds) X-ray bursts while some of them are know to show bright, high luminosity $\gamma$-ray flares~\citep[see][and the references therein]{palmer2005}. Magnetars were first discovered as Soft Gamma-ray repeaters (SGRs) in the early 80s~\citep[see][for more details]{kou87}. Since then,$\sim$30 magnetars have been discovered~\citep{olausen2014}. The usual lack of radio pulsations from most magnetars makes it difficult to detect them as they are mostly detected through episodic bursts that are observed at high energies (X-rays/$\rm \gamma-rays$). Five of these sources are known to emit pulsed radio emission, the onset of which is correlated with the onset of the outburst seen at high energies~\citep{Camilo2006, camilo2007, champion2020}. Multi-wavelength emission from magnetars gives astronomers the opportunity to study the underlying radio and X-ray emission mechanism~\citep[see][and the references therein]{rea2012} simultaneously. In recent years, magnetars have been identified as promising candidates to produce Fast Radio Bursts (FRBs)~\citep{lu2020}, millisecond duration bright radio flashes that are extragalactic in nature~\citep{petroff21}. The recent discovery of a very bright FRB from the magnetar SGR J1935+2154 contemporaneous with an X-ray burst from our very own Galaxy has shown that magnetars might be able to explain some, if not all FRBs~\citep{sgr1935, bochenek2021, li2021}. Hence, it has become increasingly important to monitor magnetars in outburst at multiple wavelengths.

Swift J1818.0-1607 is a magnetar that was discovered in early 2020 via a bright X-ray outburst. The source was detected by the \textit{Swift}-Burst Alert Telescope (BAT)~\citep{stamatikos2020}. Subsequent follow-up of the source by the \textsc{NICER} instrument revealed X-ray pulsations with a period of 1.36~seconds thus,
confirming the neutron star nature of the source~\citep{Enoto2020, Esposito2020}. Soon after, the magnetar was detected at radio wavelengths by a multitude of radio telescopes, making this only the fifth radio loud magnetar~\citep{karu2020,rajwade2020atel}. The characteristic age of the magnetar based on the spin-down rate ranges from 250--1000 years~\citep{champion2020}, making it one of the youngest magnetars known. Since the outburst, the magnetar has exhibited large variations in the spin-down behaviour. Initially, the radio emission from the magnetar also exhibited large variability with the radio pulse profile changing on a timescale of a few hours with a corresponding change in the polarization properties~\citep{champion2020}.~\cite{Lower2020} provided a detailed study of the polarization of the magnetar and showed that it exhibited a mode change whereby the radio emission abruptly changed with a corresponding change in the pulse shape pointing to a very dynamic magnetosphere of the neutron star. Recently,~\cite{torne2020} and~\cite{eie2021} have shown the evolution of the magnetar at high radio frequencies (> 2~GHz) suggesting that since the outburst, the spectral index of the radio emission has changed significantly eventually flattening with time.  While the short timescale properties of Swift J1818.0-1607 have been thoroughly investigated at radio and X-ray wavelengths~\citep{hu2020, champion2020}, long-term studies are important in order to fully understand the evolution of magnetars and gain insights into the emission mechanism. Swift~J1818.0-1607 has been regularly observed at radio and X-ray wavelengths over the span of $\sim$ 1 year since the outburst. This provides a chance to study the quasi-simultaneous evolution of the radio and X-ray emission of the magnetar since the outburst. Here, we present the evolution of the radio and X-ray emission of Swift~J1818.0-1607 over the span of 1 year as the magnetar transitions to quiescent X-ray flux levels. The paper is organised as follows: We describe the long-term radio observations of the magnetar in section~\ref{sec:obs}. Section~\ref{sec:ran} describes the analysis of the radio data while section~\ref{sec:xal} describes the analysis of the X-ray data from \textit{NICER}. We discuss our findings in section~\ref{sec:dis} and conclude in section~\ref{sec:sum}.

\section{Observations}
\label{sec:obs}
Observations were carried out using the 76-m Lovell radio telescope and the 26-m MkII telescope when the 76-m telescope was unavailable for observing, both located in Jodrell Bank in the UK. For each observation, the source was observed at a centre frequency of 1.53 GHz, with 512 MHz of bandwidth divided into 1532 channels. Complex voltage data from the telescope were converted into Stokes parameters using a ROACH-1 Field Programmable Gate Array board. Then, they were processed by the DFB backend~\citep{manchester2013}, folding the data modulo the topocentric period of the magnetar and dedispersing them at the dispersion measure (DM) of 699~pc~cm$^{-3}$. The resulting time-frequency data were folded using the best ephemeris into multiple 8-second sub-integrations and saved to disk. These data were then visually checked for frequency channels and time samples corrupted by radio frequency interference (RFI). The culprit data were flagged before saving the cleaned dynamic spectra to disk for further processing. 

At Effelsberg in Germany, Swift J1818.0$-$1607 was observed with the 100-m radio telescope. The observations were carried out at 6\,GHz using the wide-band receiver covering 4--8\,GHz. The data were recorded in \textsc{PSRFITS} search mode format~\citep{hotan2004}, with a time and frequency resolution of 131 $\mu$s and $\sim 1$ MHz, respectively  \citep[see e.g.][]{desvignes2018}. Together with each pulsar session, signals from a 1\,Hz switched noise diode were recorded for approximately 2\,min to enable polarisation calibration. The calibration was carried out for frequency-resolved differential gain and phase between the two orthogonal polarisations of the receiver. Then, a correction for a rotation measure of 1442\,rad\,m$^2$ \citep{champion2020} was implemented to account for the Faraday rotation, obtaining a polarization profile. Most of the data post-processing used the \textsc{psrchive} software package \textsc{psrchive} \citep{hotan2004}. The dates of observations used in this study are presented in Table~\ref{tab:obs_params} and the full table is available in the supplementary materials.

\section{Radio data analysis}
\label{sec:ran}
\subsection{Radio timing and spin-down}

\begin{table}
\caption{Timing parameters of Swift J1818.0-1607 for the data used in this work.}
\label{tab:timingparams}
\centering
\begin{tabular}{l l}
\hline \\ 
Parameter & Value\\ [1ex]
\hline \\
Right Ascension (J2000) [hh:mm:ss] & 18:18:00.23 \\
Declination (J2000) [dd:mm:ss] & $-$16:07:53.00 \\
DM [pc cm$^{-3}$] & 708$\pm$1 \\
Date range [MJD] & 59053.0416 -- 59426.7177 \\
Epoch of Frequency & 59280.80825 \\
F0 [Hz] & 0.732809516(29) \\
F1 [Hz s$^{-1}$] & $-1.37257(63) \times 10^{-11}$ \\
F2 [Hz s$^{-2}$] & $3.225(20) \times 10^{-19}$ \\
F3 [Hz s$^{-3}$] & $4.543(30) \times 10^{-26}$ \\
F4 [Hz s$^{-4}$] & $1.29(11) \times 10^{-33}$ \\
\hline
\end{tabular}
\end{table}


Swift J1818.0-1607 was studied at radio wavelengths in the immediate aftermath of the outburst in which it was discovered~\citep{champion2020,Esposito2020}. Similar to other magnetars~\citep[see][]{kaspi2017}, Swift~J1818.0-1607 also shows a number of distinct timing events at radio and X-ray wavelengths~\citep{champion2020,hu2020}. A phase-coherent timing solution was obtained using a combination of the Lovell and MkII data spanning $\sim1.2$ year from the start of the outburst in March 2020. Barycentred times of arrival (ToAs) for the integrated pulse profile at every epoch were generated by cross-correlating the integrated profile with a standard profile using standard \textsc{psrchive} tools for timing radio pulsars~\citep{hotan2004}. For this magnetar, we used a single standard template for all the data in spite of variations in the radio pulsed profile as is common in radio loud magnetars. The reason was that the radio profile of Swift J1818.0-1607 at 1.4~GHz always seems to have a dominant, narrow emission component that is often accompanied by wings of broad structure that appear and disappear and do not contribute to the overall precision of the TOAs. We fixed the RA and DEC value to the best known X-ray position for the entire timing taken from~\cite{blumer2020}. In this study, we focus on the timing of Swift J1818.0-1607 for a span of 1 year from July 2020 to July 2021.

\begin{figure}
	\includegraphics[scale=0.35]{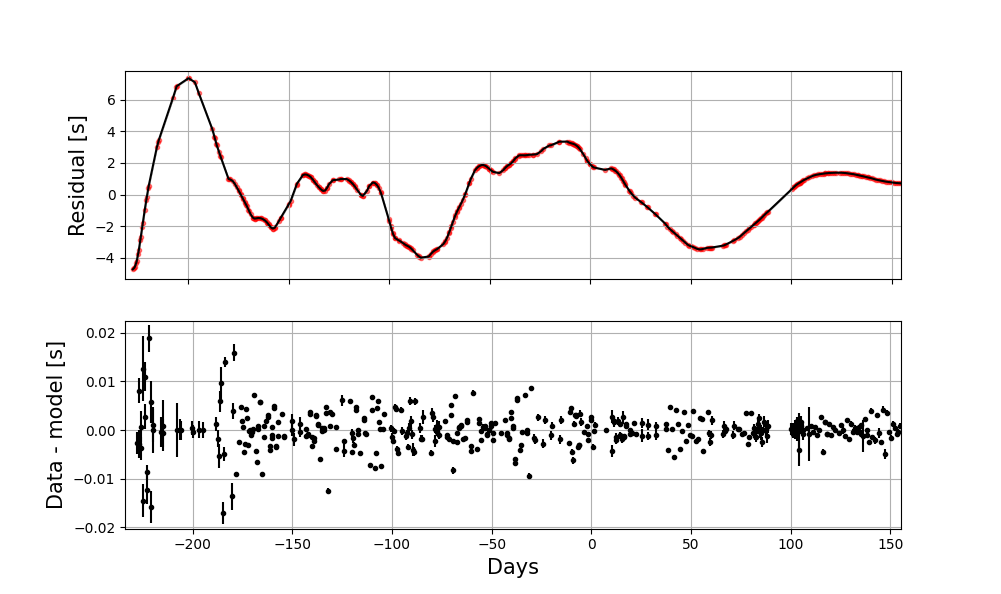} \\
    \caption{Timing residuals from the Lovell observing campaign at 1.4~GHz. The top panel shows the residuals as red dots and the black line denotes the fit to the timing residuals using a GP regression technique (see text). The individual ToA uncertainties are too small compared to the spread of the residuals to be clearly seen. The bottom panel shows the differences between the timing residuals and the GP model. The reference MJD is 59280.8.}
    \label{fig:timingres}
\end{figure}

\begin{figure*}
	\includegraphics[scale=0.46]{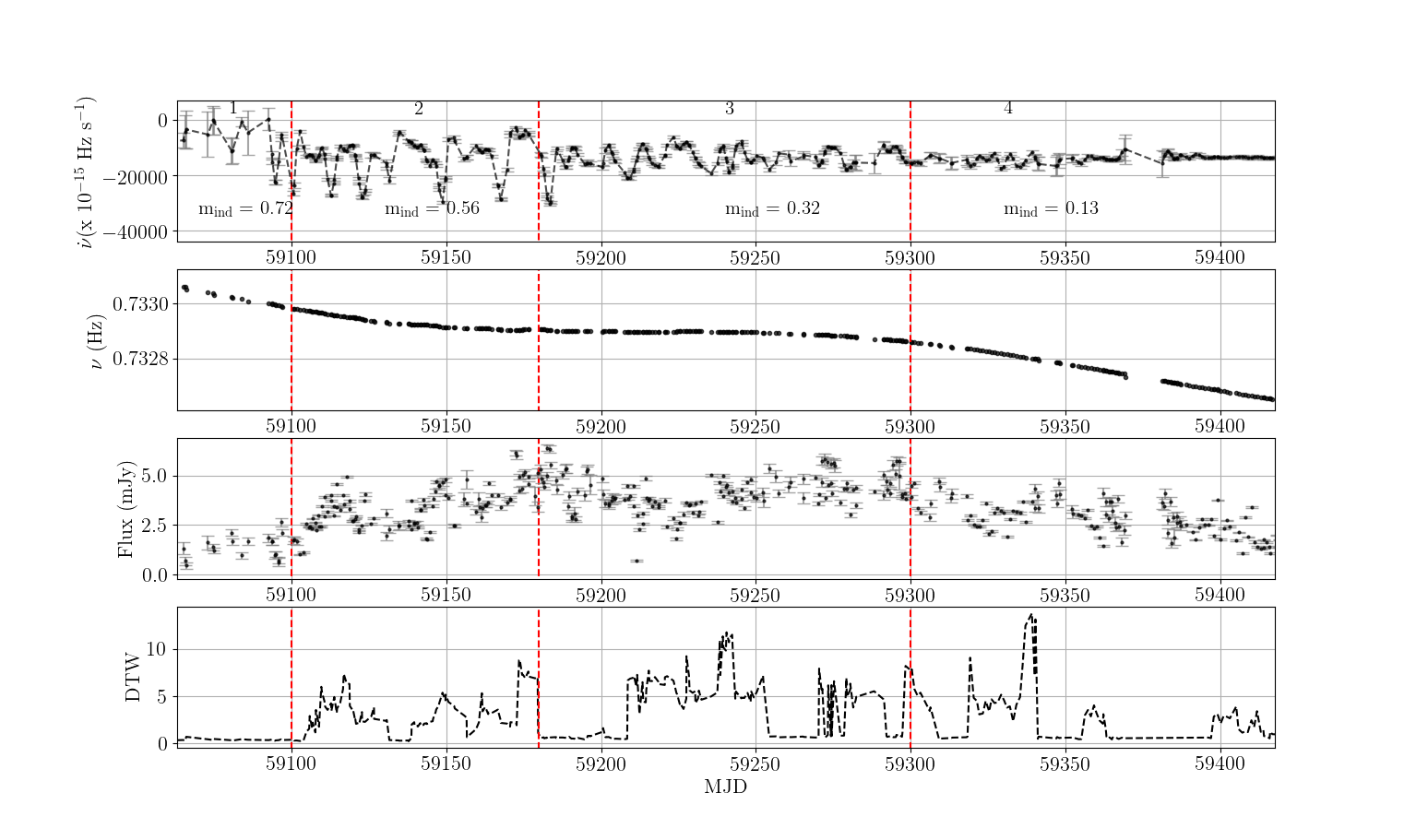}
    \caption{Top panel shows the frequency derivative as a function of time for Swift J1818.0-1607. The red vertical lines mark the approximate MJDs when the radio profile was observed to transition to a new mode from the regular observations with the Lovell telescope. The numbers mark the mode of emission as shown in figure~\ref{fig:emission}. The corresponding value of the modulation index of the frequency derivative for the span of each mode is also shown on the plot. The panels below show the spin frequency as a function of MJD, the average radio flux at 1.4~GHz as a function of MJD and the cost function corresponding to the the DTW as a function of MJD in that order (see text for more details).} 
    \label{fig:timing}
\end{figure*}

A timing model containing the spin-frequency and its first 4 derivatives was fitted to the TOAs using the \textsc{TEMPO2} pulsar timing software \citep{hobbs2006}. To look at the variation in DM, we split the Lovell data into two subbands and we realized that the DM is more or less consistent around the value of 708$\pm$1~pc~cm$^{-3}$. Hence, we decided to the keep this value of DM fixed for the analysis. It is difficult to constrain the DM in the timing analysis as we are only observing the magnetar at a single frequency with a limited bandwidth. We also used the DE200 ephemeris for this analysis. The post-fit parameters of our timing model are shown in Table~\ref{tab:timingparams} and the timing residuals (the differences between the measured TOAs and those predicted by the timing model) are shown as red points in the upper panel of Figure~\ref{fig:timingres}. It is clear that there is significant correlated structure remaining in the timing residuals, indicating that the rotational parameters are smoothly varying about some characteristic mean values.  In order to model the TOAs, we use the \textsc{scikit$-$learn} implementation of Gaussian process (GP) regression~\citep{scikit-learn} to fit a smooth, continuous function to the timing residuals~\citep[see][for full details of this process]{brook2016} . We find that the residuals are best modelled using a single squared exponential covariance function, with an additive white noise function to model the uncertainty on the residuals. The fit is applied by optimising the likelihood of the residuals, conditioned on the hyperparameters of the covariance function $\theta(\sigma^2, \lambda, \sigma^2_\mathrm{N})$, where $\sigma^2$ is the signal variance,  $\lambda$ is a characteristic scale of the modulations in the data, and $\sigma^2_\mathrm{N}$ is the noise variance.  The optimised hyperparameters were  $\sigma$=1.41 s, $\lambda$=3.15 days and $\sigma^2_\mathrm{N}$=2.35$\times$10$^{-5}$ s. The resulting model is shown as the black line in the upper panel of Figure~\ref{fig:timingres}. The lower panel shows the difference between the GP model and the timing residuals. The larger residuals at $\sim$-220 and -180 are due to very faint detection of the magnetar due to reduced sensitivity and RFI at the telescope. We then use the optimised hyperparameters to compute the first and the second derivative of the GP model of the timing residuals, in line with \cite{brook2016}, resulting in values of the spin-frequency (F0) and spin-frequency derivative (F1) respectively. Figure~\ref{fig:timing} shows the evolution (from top to bottom) of F1, F0, radio flux at 1.4~GHz and the Dynamic Time Warp (DTW) metric with time (explained in  section~\ref{ssec:radevo}).

\begin{figure}
	\includegraphics[scale=0.48]{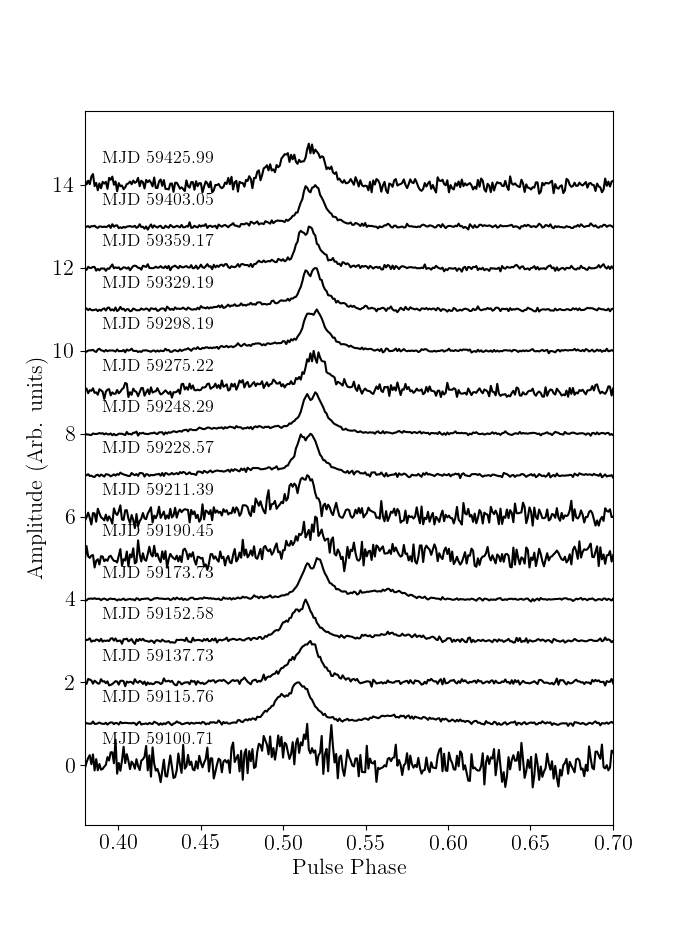}
    \caption{Folded average radio profile of Swift~J1818.0-1607 at 1.4 GHz from the Lovell/MkII observations. The MJD corresponds to the date of observation for which the folded profile is generated.}
    \label{fig:Lovellprofs}
\end{figure}

~\cite{champion2020} report on a number of 'events' that caused a big change in the spin-frequency of the magnetar. Figure~\ref{fig:timing} shows the evolution of these parameters beyond the MJD range studied in~\cite{champion2020}. The frequency derivative of the magnetar is highly variable and seems to fluctuate around a mean value of $-$1.37$\times$10$^{-11}$Hz~s$^{-1}$. While the fluctuations start out as being fairly large in magnitude ($\Delta {\rm F1}\sim$10$^{-11}$~Hz~s$^{-1}$), as time goes on, the magnitude of the variation keeps decreasing gradually. By MJD 59350, the frequency derivative seems to converge to the mean value. To quantify this behaviour, we defined a modulation index $\rm m_{\rm ind}$ as the ratio of the standard deviation of the frequency derivative and the mean frequency derivative. Using this metric, we were able to identify four regions in the plot (separated by red dashed lines) that were determined based on four distinct values of the modulation index, as seen in the top panel of figure~\ref{fig:timing}. 

We also computed the radio flux density of Swift J1818.0-1607 for every epoch in our dataset. Approximately daily observations were carried out of the Crab pulsar, which naturally included the Crab Nebula. The difference in the instrumental output levels between observations of the Crab Nebula and a nearby "cold sky" reference position allowed determination of the flux density scale of the receiver output, and hence of any pulse profile.~\cite{vinyaikin2007} have shown that the radio flux density of the Crab Nebula only changes by 0.2$\%$ every year and hence, is stable enough to be used as a calibrator. This allowed calculation of the time-averaged mean flux density. The same procedure was carried out for the observations using both the 76-m Lovell telescope and the 26-m MkII telescope and the resultant time sequence is also shown in Figure~\ref{fig:timing}.

\subsection{Radio pulse profile evolution}
\label{ssec:radevo}
\begin{figure*}
	\includegraphics[scale=0.53]{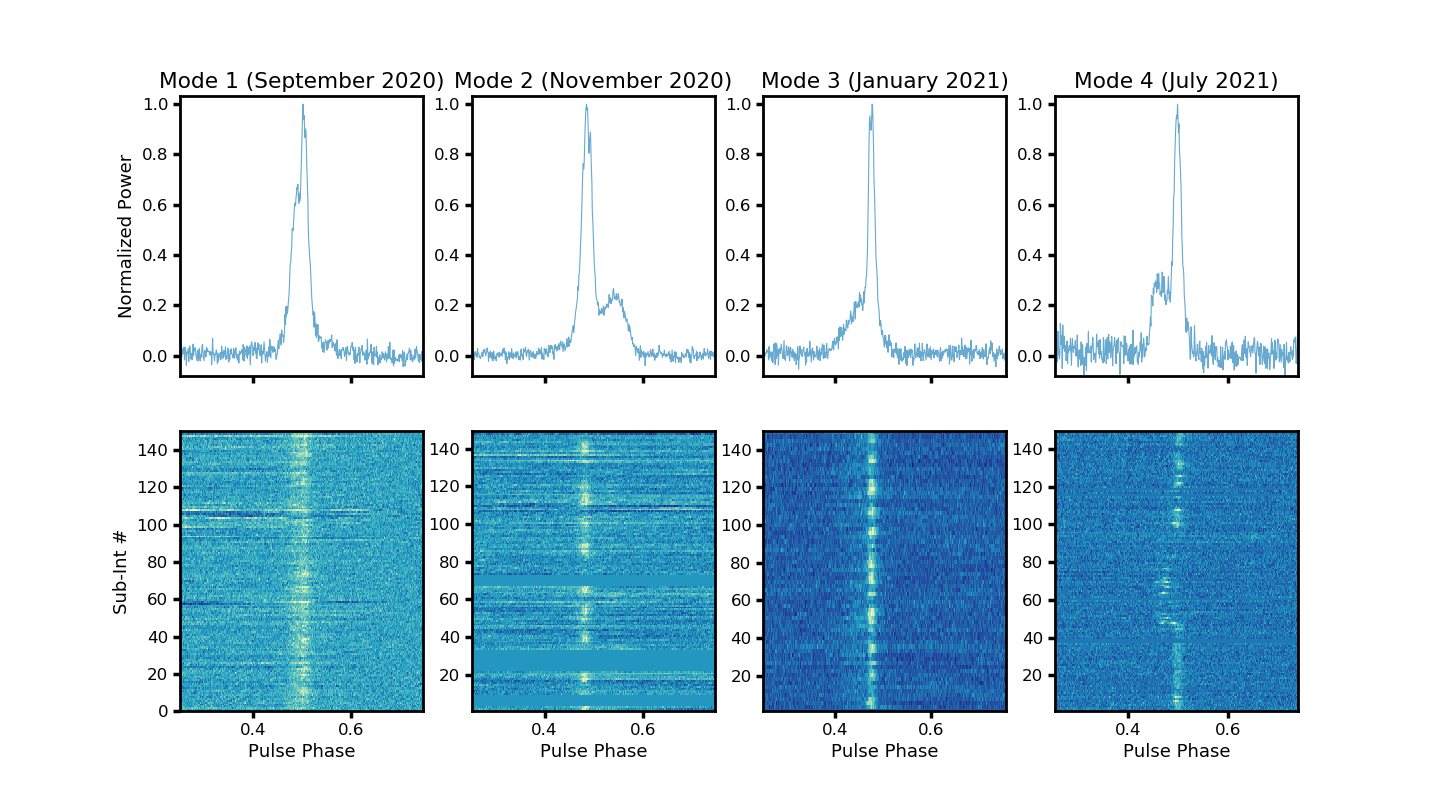}
    \caption{The four different radio emission modes of Swift J1818.0-1607 that we identified during this campaign. The bottom panel shows 8-second sub-integration timeseries for each emission mode at 1.4 GHz observed with the Lovell telescope. The top panel shows the resulting average profile formed from the timeseries shown in the bottom panel.} 
    \label{fig:emission}
\end{figure*}

\begin{figure*}
	\includegraphics[scale=0.52]{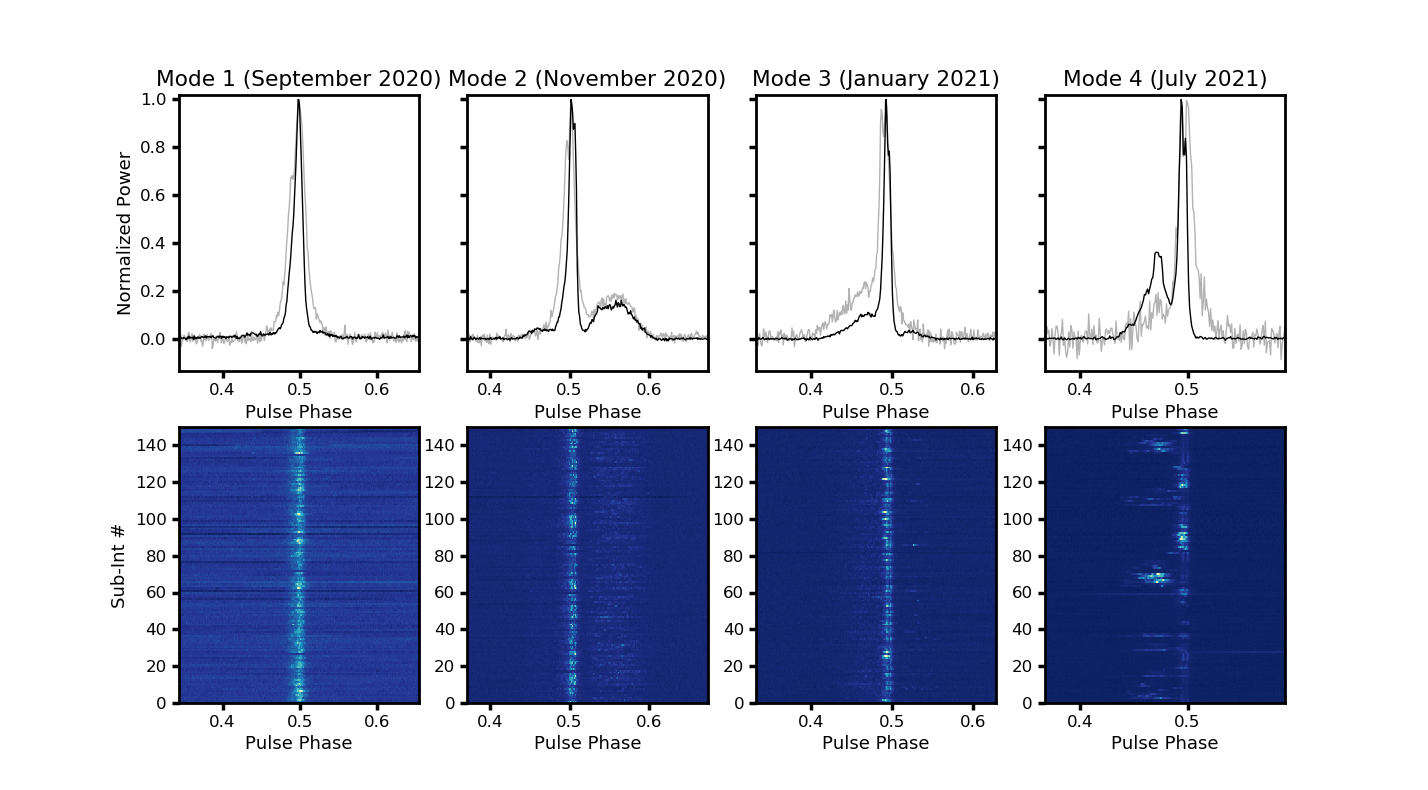}
    \caption{A plot that shows the average pulse profile of Swift J1818.0-1607 during four different modes (see text for more details). The grey profiles are observed at 1.4~GHz using the Lovell telescope while the black profiles are at 6~GHz with the Effelsberg telescope. The bottom panel shows 10-second sub-integration timeseries for each emission mode at 6~GHz observed with the Effelsberg telescope.} 
    \label{fig:freqprof}
\end{figure*}

Here, we present a detailed analysis of the profile evolution of Swift~J1818.0-1607 over the span of one year since the outburst. The magnetar has exhibited large variability in the radio emission with the appearance and disappearance of new components in the average pulse profile. Figure~\ref{fig:Lovellprofs} shows the evolution of the pulse profile as a function of MJD. The profile went from being a two narrow component profile to one having a bright leading component and a new, wide trailing component only to return to a narrow two component profile as before. Similar changes were seen in the pulse profile at 6~GHz albeit the profile components were narrower in width and hence, easily distinguishable compared to the 1.4~GHz profiles (see Figure~\ref{fig:freqprof}). It is known that there is a strong correlation between radio pulse profile changes and spin-down variations in a subset of mode-changing pulsars~\citep{lyne2010}. These pulsars typically tend to show large timing noise in the timing residuals. It is also known that magnetars are extremely noisy timers~\citep{Caleb2021} and show large variability in the spin-down rate along with significant changes in radio pulse shape~\citep{camilo2018}. Over the course of the regular radio monitoring of Swift J1818.0-1607, we identify four different emission modes (see Figure~\ref{fig:emission}). The magnetar switched from a mode that showed steady radio emission with minimal variability over short timescales (mode 1) to a mode in which a trailing component emerged in the radio profile with quasi-periodic flux modulation in the leading/bright component (mode 2). The next transition resulted in the emergence of a weaker leading component (mode 3) with continued quasi-periodic modulation of the flux which was at a shorter period compared to what was seen in mode 2. The final transition occurred in April of 2021 when the magnetar started to show sporadic nulls and offsets in the emission phase over a timescale of a few minutes very similar to the so-called `swooshes' seen in PSR~B0919+06 and PSR B1859+07~\citep{rankin2006,perera2015,perera2016, rajwade2021}. 

We note that while the folded radio profile of the magnetar remained stable over a duration of months, subtle changes in the radio emission and pulse shapes were seen on a timescale of minutes to hours. In order to check whether the change in the spin-down rate of Swift~J1818.0-1607 was correlated with the short time-scale changes in the radio emission, we quantify the change in the average pulse shape for every epoch. We use the Dynamic Time Warping algorithm to compares the change in the shape of successive radio pulse profiles~\citep[see][for more details]{keogh2005}. For two pulse profiles from the $i^{\rm th}$ and $(i+1)^{\rm th}$ epoch, $P_{i}$ and $P_{i+1}$, we define the cost function,
\begin{equation}
    \theta_{i} = |P_{{\rm i},k} - P_{{\rm i+1},k+\delta}|,
\end{equation}
where, $k$ defines the time sample and $\delta$ defines a time sample such that $\delta \geq k$. Then, the change between $P_{\rm i}$ and $P_{\rm i+1}$ is the minimum of the cost function across the entire folded profile. Hence, the dynamic time warping metric,
\begin{equation}
    \rm DTW = \min\sum_{i=0}^{i=N} \theta_{i},
\end{equation}
where N is the total number of time bins across the pulse profile.
Larger values of $\rm DTW$ indicates a larger change in the profile compared to the previous one. We computed the DTW for all consecutive pairs of folded profiles. Bottom panel of Figure~\ref{fig:timing} shows the value of the metric as a function of MJD. One can clearly see that the DTW varies between successive epochs suggesting subtle changes in the folded pulse profile on a day timescale even within a single emission mode. On the other hand, no obvious correlation can be seen between transition of emission modes and the DTW. We do note that the transition from mode 3 to 4 does correspond to a large value of the DTW (MJD 59300) although that might be coincidence due to the lack of similar trend during other mode transitions.

\section{X-ray observations and data analysis}
\label{sec:xal}
Since the onset of the outburst, Swift~J1818.0-1607 was regularly observed using the Neutron star Interior Composition ExploreR (NICER) instrument onboard the International Space Station. NICER has monitored the source over a span of 6 months since the outburst resulting in a high cadence X-ray dataset that can be used for studying the long term X-ray evolution of the magnetar to draw comparisons with the radio evolution. Here, we study the publicly available data since March of 2020 from the HEASARC NICER data archive~\footnote{\url{https://heasarc.gsfc.nasa.gov/db-perl/W3Browse/w3table.pl?tablehead=name\%3Dnicermastr\&Action=More+Options}}. Since we study the long-term evolution, we use one observation from every month, thus limiting the study to variability over month-long timescales. We note that the results from the X-ray timing analysis of this magnetar are being compiled as a separate paper (Bansal et al., in prep.)

For calibration and filtering of the data, we used HEASOFT version 6.29 and NICERDAS version 2021-08-31V008c. We applied the standard filtering criteria (excluding events acquired during times of South Atlantic Anomaly passage and with pointing offsets greater than 54$\arcsec$; including data obtained with Earth elevation angles greater than 30° above the dark limb and 40° above the bright limb). We also ignored all the spurious photons as a result of optical loading due to the Sun. Using these filters, we used the standard NICERDAS suite \textsc{nicerl2} to calibrate, filter and extract good time intervals (GTIs) for all our observations. Because NICER consists of non-imaging detectors, the NICER team has developed a model for the expected X-ray background. Using the \textsc{nibackgen3C50} tool provided by the NICER team~\footnote{\url{https://heasarc.gsfc.nasa.gov/docs/nicer/tools/nicer_bkg_est_tools.html}}, we estimated the background spectrum for each observation segment separately.

\begin{table*}
\caption{The 0.5--12~keV spectral parameters for Swift~J1818.0-1607 for a 1BB model. The X-ray flux corresponds to the unabsorbed flux. We note that the estimated radius is at infinite distance (does not account for the gravitational redshift of the surface emission).}
\label{tab:specparams}
\centering
\begin{tabular}{l l l l l }
\hline \\ 
Date  &$F_{\rm X, BB}$ (0.5--12~keV)&$kT_{\rm warm}$  & Radius \\ [1ex]
(MJD) & ($\times$10$^{-11}$ ergs~cm$^{-2}$~s$^{-1}$) &(keV)& (km) \\ [1ex]
\hline \\
58927.39199074 & 1.64$\pm$0.04 & 1.05$\pm$0.03 & 2.7 $\pm$0.3 \\
58939.52222222 & 1.40$\pm$0.20 & 1.06$\pm$0.04 & 2.2 $\pm$0.2 \\
58962.31253472 & 1.25$\pm$0.05 & 1.11$\pm$0.05 & 1.6$\pm$0.1 \\
58972.25486111 & 1.31$\pm$0.09 & 1.14$\pm$0.03 & 1.4$\pm$0.2  \\
59011.1937500 & 0.65$\pm$0.03 & 0.97$\pm$0.05 & 1.6$\pm$0.3 \\
59047.27846065 & 0.69$\pm$0.01 & 1.05$\pm$0.04 & 1.1$\pm$0.3 \\
59066.25134259 & 0.68$\pm$0.08 & 1.09$\pm$0.08 & 1.0$\pm$0.2 \\
59109.11567130 & 0.75$\pm$0.06 & 0.90$\pm$0.06 & 1.7$\pm$0.4 \\
59129.67307870 & 0.59$\pm$0.09 & 1.04$\pm$0.02 & 1.0$\pm$0.3 \\
59159.094756948 & 0.45$\pm$0.05 & 0.99$\pm$0.04 & 1.0$\pm$0.2 \\
59296.170173611 & 0.21$\pm$0.01 & 0.87$\pm$0.05 & 0.9$\pm$0.4 \\
\hline
\end{tabular}
\end{table*}

\subsection{X-ray spectroscopy}
To study the long-term spectral evolution of Swift~J1818.0-1607, each observation segment shown in Table~\ref{tab:specparams} was analyzed separately. Previous studies~\citep{blumer2020} have shown that Swift~J1818.0-1607 is a highly absorbed source hence the X-ray background dominates the super-soft region of the X-ray spectrum. We assume that the X-ray emission is dominated by the thermal X-rays coming from an hot-spot on the surface of the neutron star~\citep{hu2020}. As the background dominates the spectrum above 7.5~keV and below 2.5~keV, we only fit the spectra in the 2.5--7.5~keV range. For the spectral fitting, we used the \textsc{tbabs} x \textsc{bbodyrad} model in \textsc{xspec} \citep{Arnaud} assuming interstellar medium (ISM) abundances \citep{Wilms}. The \textsc{bbodyrad} model directly provides a way to compute the area of the hot-spot through the normalization factor. The spectra were binned in such a way that each spectral channel had at least 25 counts. This was done to improve the S/N ratio per spectral channel for properly fitting the data. Then, we jointly fit the data from all the epochs chosen for this analysis using the above mentioned model and fixing the value of nH to 1.13$\times$10$^{23}$~cm$^{-2}$ from~\cite{Esposito2020}. We assume that the uncertainty on the background is Gaussian while the photons from the magnetar are Poisson distributed. Hence, we used the \textsc{pgstat} statistic in order to measure the goodness of fit. The final best-fit parameters for all the spectra considered in this work are shown in Table~\ref{tab:specparams}. For the joint fit, we obtained a reduced $\chi^{2}$ of 1.09 with 876 degrees of freedom.

Figure~\ref{fig:xevo} shows the evolution of the different spectral parameters as a function of time for Swift J1818.0-1607. One can see that initially, for approximately 4-5 months after the onset of the outburst, the temperature of the hot-spot remained fairly constant while the area of the hot-spot kept reducing resulting in a decrease in the total 0.5--10~keV flux. This is consistent with the X-ray flux evolution seen for other magnetars after an outburst~\citep{zelati2017}.

\section{Discussion}
\label{sec:dis}
Radio-loud magnetars provide an avenue to study the evolution of highly magnetized neutron stars that is typically not straightforward at high energies once the X-ray flux has reached quiescent levels. The only way to characterize the spin evolution of radio-quiet magnetars is during the course of an outburst that can last for anywhere between a few months to an year~\citep{kaspi2017} after which one loses the spin evolution history of the source until the next outburst. On the other hand, radio emission from magnetars tends to persist for much longer which means that they can be timed precisely in order to give us insight into the dynamic magnetosphere of magnetars in the aftermath of an outburst. Similar to other magnetars, Swift J1818.0-1607 showed extreme variability in the torque (spin-frequency derivative) in the initial days after the outburst~\citep{champion2020} before switching to a steady decrease in the spin-frequency derivative as observed for other radio-loud magnetars~\citep[for e.g.][]{camilo2018, levin2018}. Typically, erratic variations in torque are observed at the end of this steady decrease phase which can last for months. While the erratic phase has been commonly observed in other magnetars like XTE J1810-197 and PSR J1622-4950~\citep{Caleb2021, camilo2018}, a more systematic spin-down behaviour has been observed for Swift J1818.0-1607. Figure~\ref{fig:timingres} shows the frequency derivative evolution for Swift J1818.0-1607 over the span of one year since the outburst. The value of F1 seems to converge to a value of $-$1.37$\times$10$^{-11}$~Hz~s$^{-1}$. While the behaviour of the frequency derivative observed is not periodic, it is reminiscent of a damped harmonic oscillator. A few theoretical models have been proposed where magnetars can undergo toroidal oscillations triggered by an outburst~\citep[see][and the references therein]{morozova2011}. These oscillations can then result in cyclical variations in the torque on the neutron star to get the observed behaviour of F1. Assuming that the F1 converges to the long-term value of F1, we computed the characteristic age ($\tau_{c}$) of the magnetar and we obtain $\tau_{c} \simeq$ 860 years. This value is almost a factor of 4 larger than $\tau_{c}$ reported in~\cite{Esposito2020} but still might be one of the youngest magnetars ever known based on the characteristic ages of other magnetars~\citep{olausen2014}.

 We already reported that the radio emission from Swift J1818.0-1607 can be grouped into four emission modes. From the daily radio monitoring, we observe that the transition between different modes is gradual and we can visually identify the approximate MJD of the onset of transition between different emission modes. Interestingly, these MJDs for transition coincide with the change in the modulation index of the spin frequency derivative (see top panel of figure~\ref{fig:timing}). This suggests a correlation between the change of the spin-down state and the radio emission of the magnetar. This type of behaviour is similar to what is seen in some mode-changing radio pulsars where the spin-frequency switches between two or more states with a corresponding change in the radiative properties of the neutron star~\citep{lyne2010}. The bottom panel of  figure~\ref{fig:timing} shows that while there might not be a direct correlation between a change in F1 and the radio emission profile, the radiative properties of the magnetar are correlated with macroscopic changes in F1 that are characterized by the modulation index. From Figure~\ref{fig:timing}, it is not obvious that the change in the modulation index of the spin-down is directly correlated to the change in the DTW metric before and after the transition in the mode. This suggests that the transition between the modes is gradual rather than abrupt, unlike what was seen in the early stages of the outburst~\citep{champion2020}. The Effelsberg radio telescope observations at 6~GHz helped us compare the profile evolution with radio frequency (see Figure~\ref{fig:freqprof}). The figure shows that the profiles at 6~GHz are narrower in width compared to the ones at 1.4~GHz conforming with what is expected from the radius to frequency mapping for canonical pulsars~\citep{cordes78}. We do acknowledge that we cannot rule out the possibility that the increase in width at lower frequency can be attributed to propagation effects in the Inter-Stellar Medium like  scintillation and scattering~\citep[for e.g. see]{rickett1977}. Assuming radius to frequency mapping to be true for Swift J1818.0-1607, the radio emission in this magnetar is most likely coming from the region just above the polar cap rather than close to the light cylinder~\citep[see][and references therein]{kramer2007} which is also corroborated by the sweep of the polarization angle of the radio emission~\citep{Lower2020}. This is unlike what is seen for other radio loud magnetars that show complex pulse shapes and polarization angle sweeps~\citep[for e.g.][]{camilo2007,camilo2018,Caleb2021}.  Intriguingly, the 6~GHz profile during mode 2 shows a weak leading component that is not present at all in the 1.4~GHz profile. Similarly, mode 3 profile at 6~GHz shows a weak trailing component but one can argue that the same component is not visible in the 1.4~GHz data due to the lack of sensitivity. Overall, the profiles at higher frequencies exhibit emission components that are well separated compared to the ones at 1.4~GHz. 

\begin{figure*}
\centering
	\includegraphics[scale=0.5]{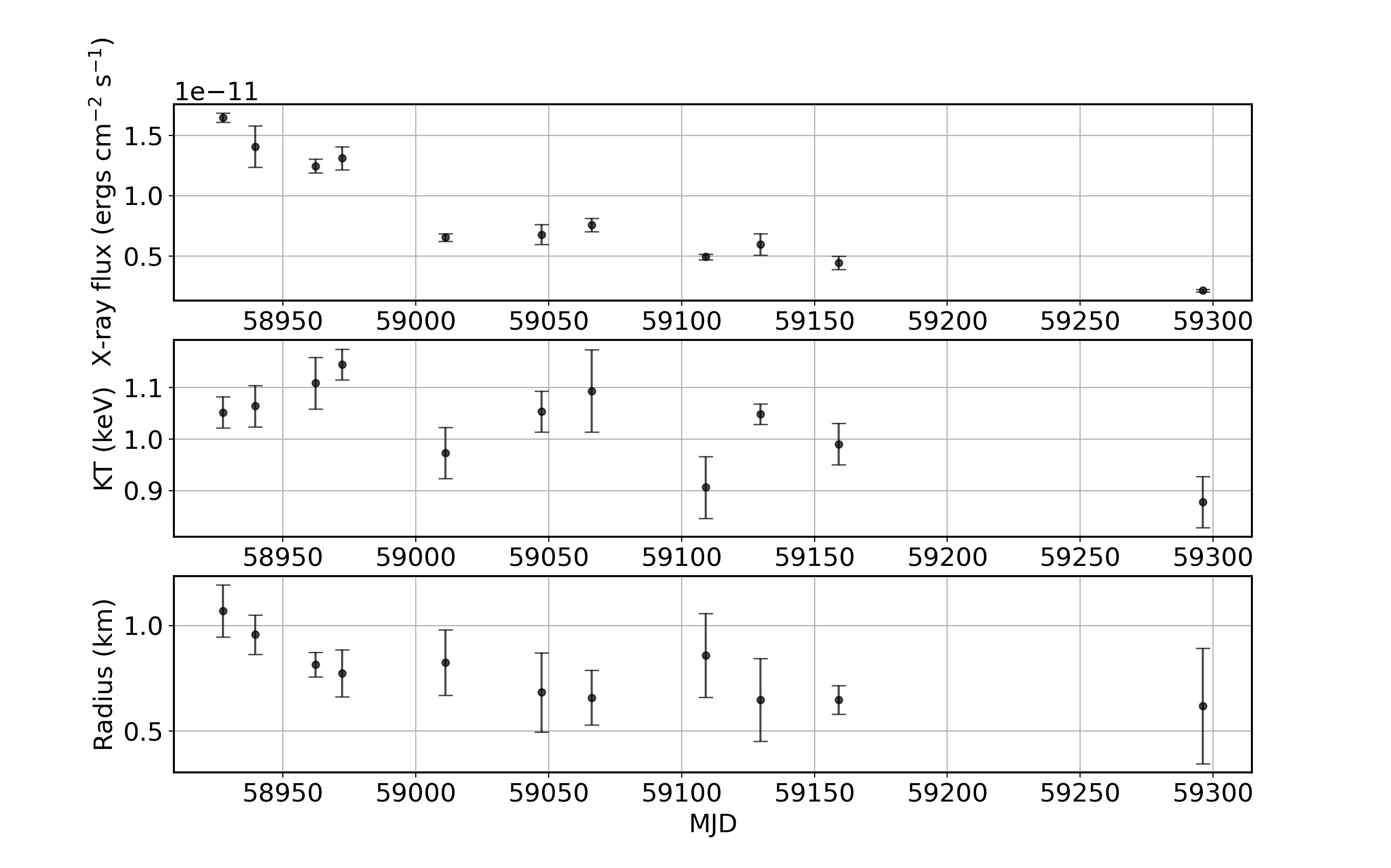}
    \caption{Evolution of various derived parameters from the 
    spectral fitting of the NICER data of Swift J1818.0-1607. The top 
    panel shows the 0.5--12~keV X-ray flux, the middle panel shows 
    the blackbody temperature and the bottom panel shows the radius of 
    the emitting region. For obtaining the radius, we assumed a distance 
    of 6.5~kpc from Blumer et al. (2020).}
    \label{fig:xevo}
\end{figure*}

Figure~\ref{fig:xevo} shows the evolution of the X-ray flux and the blackbody component coming from the hot-spot on the surface of the neutron star. The 0.5--12~keV X-ray flux shows an exponentially decreasing trend. In order to quantify this, we fit the flux with an exponential function to obtain a time constant ($\tau$) of 187.7$\pm$9.6 days with a reduced $\chi^{2}$ of 2.15 with 7 degrees of freedom. The value of $\tau$ obtained is consistent with the timescale observed for other magnetars~\citep{bernardini2009}. Such a decrease can be attributed to a blackbody component from the surface of the neutron star with a fairly constant temperature (~1.1~keV) but a decreasing area of the emission region. Typically, magnetars show evidence of cooling down of the blackbody component itself anywhere between a few months to few years after the onset of the outburst. In the case of Swift J1818.0-1607, the blackbody component exhibits no indication of cooling down in the 1 year since the outburst. This is different from another radio loud magnetar XTE~J1810-197 where the two hot-spots detected by NICER started to cool down approximately 6 months after the outburst~\citep{Caleb2021}. Unlike XTE J1810-197, Swift J1818.0-1607 is a heavily absorbed source owing to the large value of nH along the line of sight hence the source is barely detected in a 1~ks observation by NICER in March 2021. Assuming that the blackbody component has cooled down below 1~keV since then, it will be impossible to detect the source with NICER on account of the high background below 2~keV. More sensitive instruments like \textit{Chandra} and \textit{XMM-Newton} would be needed to observe the X-ray emission to determine when the blackbody starts to cool down. Comparing Figures~\ref{fig:timing} and~\ref{fig:xevo}, we can conclude that there is no correlation between the evolution of the radio and the X-ray flux. The X-ray flux continues to decrease exponentially as the radio flux keeps increasing. This lack of correlation between the radio and X-ray flux is similar to what is seen for other radio-loud magnetars~\citep{camilo2018, Caleb2021} suggesting that while the outburst creates the required plasma in the magnetosphere to kick start the mechanism producing coherent radio emission, the processes that create the X-ray and radio emission are not physically tied to each other such that small changes in one do not affect the other.

\section{Summary}
\label{sec:sum}
In summary, we have presented the results of a long term radio and X-ray monitoring campaign of a young magnetar Swift~J1818.0-1607. The radio campaign shows that the radio emission of the magnetar evolves over time and can be characterized by distinct modes defined by distinct radio profile shapes and emission. At the same time, the long term timing of the magnetar shows similar modes that are characterized by the change in the modulation index of the spin-frequency derivative about a nominal value of $-$1.37$\times$10$^{-11}$~Hz~s$^{-1}$. The time of the transition between two radio modes is correlated to the time of transition in the mode of the spin frequency derivative of the magnetar indicating a change in physical conditions of the magnetosphere  similar to what is seen in other mode-changing canonical pulsars. This further consolidates the physical link between canonical pulsars and radio-loud magnetars and suggests that the radio emission in them are different manifestations of the same emission mechanism~\citep{kramer2007}. We also show that in this time, the X-ray flux evolution is completely uncorrelated to the radio emission and follows an exponential decay that can be attributed to a thermal hot-spot on the surface of the star with a constant temperature and decreasing surface area. Our study shows that the radio emission in magnetars can be quite similar to the emission seen from radio pulsars and Swift J1818.0-1607 might be the link that connects the two sub-populations of neutron stars. It also demonstrates the importance of long-term monitoring of these sources at multiple wavelengths in order unravel the underlying emission mechanism of neutron stars.

\section*{Acknowledgements}
The authors would like to thank the anonymous referee for constructive comments that vastly improved the manuscript. KMR, BWS and MS acknowledge funding from the European Research Council (ERC) under the European Union's Horizon 2020 research and innovation programme (grant agreement No. 694745). KL, GD, RK acknowledge the financial support by the European Research Council for the ERC Synergy Grant BlackHoleCam under contract no. 610058. Pulsar research at Jodrell Bank Centre for Astrophysics and Jodrell Bank Observatory is supported by a consolidated grant from the UK Science and Technology Facilities Council (STFC). This publication is based on observations with the 100-m telescope of the MPIfR (Max-Planck-Institut für Radioastronomie) at Effelsberg. 
\section*{Data Availability}

The data used in this manuscript will be made available to others upon reasonable request to the authors.



\bibliographystyle{mnras}
\bibliography{main} 

\begin{thebibliography}{}
\makeatletter
\relax
\def\mn@urlcharsother{\let\do\@makeother \do\$\do\&\do\#\do\^\do\_\do\%\do\~}
\def\mn@doi{\begingroup\mn@urlcharsother \@ifnextchar [ {\mn@doi@}
  {\mn@doi@[]}}
\def\mn@doi@[#1]#2{\def\@tempa{#1}\ifx\@tempa\@empty \href
  {http://dx.doi.org/#2} {doi:#2}\else \href {http://dx.doi.org/#2} {#1}\fi
  \endgroup}
\def\mn@eprint#1#2{\mn@eprint@#1:#2::\@nil}
\def\mn@eprint@arXiv#1{\href {http://arxiv.org/abs/#1} {{\tt arXiv:#1}}}
\def\mn@eprint@dblp#1{\href {http://dblp.uni-trier.de/rec/bibtex/#1.xml}
  {dblp:#1}}
\def\mn@eprint@#1:#2:#3:#4\@nil{\def\@tempa {#1}\def\@tempb {#2}\def\@tempc
  {#3}\ifx \@tempc \@empty \let \@tempc \@tempb \let \@tempb \@tempa \fi \ifx
  \@tempb \@empty \def\@tempb {arXiv}\fi \@ifundefined
  {mn@eprint@\@tempb}{\@tempb:\@tempc}{\expandafter \expandafter \csname
  mn@eprint@\@tempb\endcsname \expandafter{\@tempc}}}

\bibitem[\protect\citeauthoryear{{Arnaud}, {Dorman}  \& {Gordon}}{{Arnaud}
  et~al.}{1999}]{Arnaud}
{Arnaud} K.,  {Dorman} B.,   {Gordon} C.,  1999, {XSPEC: An X-ray spectral
  fitting package} (\mn@eprint {ascl} {9910.005})

\bibitem[\protect\citeauthoryear{{Bernardini} et~al.,}{{Bernardini}
  et~al.}{2009}]{bernardini2009}
{Bernardini} F.,  et~al., 2009, \mn@doi [\aap] {10.1051/0004-6361/200810779},
  \href {https://ui.adsabs.harvard.edu/abs/2009A&A...498..195B} {498, 195}

\bibitem[\protect\citeauthoryear{{Blumer} \& {Safi-Harb}}{{Blumer} \&
  {Safi-Harb}}{2020}]{blumer2020}
{Blumer} H.,  {Safi-Harb} S.,  2020, \mn@doi [\apjl]
  {10.3847/2041-8213/abc6a2}, \href
  {https://ui.adsabs.harvard.edu/abs/2020ApJ...904L..19B} {904, L19}

\bibitem[\protect\citeauthoryear{{Bochenek}, {Ravi}, {Belov}, {Hallinan},
  {Kocz}, {Kulkarni}  \& {McKenna}}{{Bochenek} et~al.}{2020}]{bochenek2021}
{Bochenek} C.~D.,  {Ravi} V.,  {Belov} K.~V.,  {Hallinan} G.,  {Kocz} J.,
  {Kulkarni} S.~R.,   {McKenna} D.~L.,  2020, \mn@doi [\nat]
  {10.1038/s41586-020-2872-x}, \href
  {https://ui.adsabs.harvard.edu/abs/2020Natur.587...59B} {587, 59}

\bibitem[\protect\citeauthoryear{{Brook}, {Karastergiou}, {Johnston}, {Kerr},
  {Shannon}  \& {Roberts}}{{Brook} et~al.}{2016}]{brook2016}
{Brook} P.~R.,  {Karastergiou} A.,  {Johnston} S.,  {Kerr} M.,  {Shannon}
  R.~M.,   {Roberts} S.~J.,  2016, \mn@doi [\mnras] {10.1093/mnras/stv2715},
  \href {https://ui.adsabs.harvard.edu/abs/2016MNRAS.456.1374B} {456, 1374}

\bibitem[\protect\citeauthoryear{{CHIME/FRB Collaboration} et~al.,}{{CHIME/FRB
  Collaboration} et~al.}{2020}]{sgr1935}
{CHIME/FRB Collaboration} et~al., 2020, \mn@doi [\nat]
  {10.1038/s41586-020-2863-y}, \href
  {https://ui.adsabs.harvard.edu/abs/2020Natur.587...54C} {587, 54}

\bibitem[\protect\citeauthoryear{{Caleb} et~al.,}{{Caleb}
  et~al.}{2021}]{Caleb2021}
{Caleb} M.,  et~al., 2021, \mn@doi [\mnras] {10.1093/mnras/stab3223}, \href
  {https://ui.adsabs.harvard.edu/abs/2021MNRAS.tmp.2944C} {}

\bibitem[\protect\citeauthoryear{{Camilo}, {Ransom}, {Halpern}, {Reynolds},
  {Helfand}, {Zimmerman}  \& {Sarkissian}}{{Camilo} et~al.}{2006}]{Camilo2006}
{Camilo} F.,  {Ransom} S.~M.,  {Halpern} J.~P.,  {Reynolds} J.,  {Helfand}
  D.~J.,  {Zimmerman} N.,   {Sarkissian} J.,  2006, \mn@doi [\nat]
  {10.1038/nature04986}, \href
  {https://ui.adsabs.harvard.edu/abs/2006Natur.442..892C} {442, 892}

\bibitem[\protect\citeauthoryear{{Camilo}, {Ransom}, {Halpern}  \&
  {Reynolds}}{{Camilo} et~al.}{2007}]{camilo2007}
{Camilo} F.,  {Ransom} S.~M.,  {Halpern} J.~P.,   {Reynolds} J.,  2007, \mn@doi
  [\apjl] {10.1086/521826}, \href
  {https://ui.adsabs.harvard.edu/abs/2007ApJ...666L..93C} {666, L93}

\bibitem[\protect\citeauthoryear{{Camilo} et~al.,}{{Camilo}
  et~al.}{2018}]{camilo2018}
{Camilo} F.,  et~al., 2018, \mn@doi [\apj] {10.3847/1538-4357/aab35a}, \href
  {https://ui.adsabs.harvard.edu/abs/2018ApJ...856..180C} {856, 180}

\bibitem[\protect\citeauthoryear{{Champion} et~al.,}{{Champion}
  et~al.}{2020}]{champion2020}
{Champion} D.,  et~al., 2020, \mn@doi [\mnras] {10.1093/mnras/staa2764}, \href
  {https://ui.adsabs.harvard.edu/abs/2020MNRAS.498.6044C} {498, 6044}

\bibitem[\protect\citeauthoryear{{Cordes}}{{Cordes}}{1978}]{cordes78}
{Cordes} J.~M.,  1978, \mn@doi [\apj] {10.1086/156218}, \href
  {https://ui.adsabs.harvard.edu/abs/1978ApJ...222.1006C} {222, 1006}

\bibitem[\protect\citeauthoryear{{Coti Zelati}, {Rea}, {Pons}, {Campana}  \&
  {Esposito}}{{Coti Zelati} et~al.}{2017}]{zelati2017}
{Coti Zelati} F.,  {Rea} N.,  {Pons} J.~A.,  {Campana} S.,   {Esposito} P.,
  2017, in Journal of Physics Conference Series. p. 012022,
  \mn@doi{10.1088/1742-6596/932/1/012022}

\bibitem[\protect\citeauthoryear{{Desvignes} et~al.,}{{Desvignes}
  et~al.}{2018}]{desvignes2018}
{Desvignes} G.,  et~al., 2018, \mn@doi [\apjl] {10.3847/2041-8213/aaa2f8},
  \href {https://ui.adsabs.harvard.edu/abs/2018ApJ...852L..12D} {852, L12}

\bibitem[\protect\citeauthoryear{{Eie} et~al.,}{{Eie} et~al.}{2021}]{eie2021}
{Eie} S.,  et~al., 2021, arXiv e-prints, \href
  {https://ui.adsabs.harvard.edu/abs/2021arXiv210911739E} {p. arXiv:2109.11739}

\bibitem[\protect\citeauthoryear{{Enoto} et~al.,}{{Enoto}
  et~al.}{2020}]{Enoto2020}
{Enoto} T.,  et~al., 2020, The Astronomer's Telegram, \href
  {https://ui.adsabs.harvard.edu/abs/2020ATel13551....1E} {13551, 1}

\bibitem[\protect\citeauthoryear{{Esposito} et~al.,}{{Esposito}
  et~al.}{2020}]{Esposito2020}
{Esposito} P.,  et~al., 2020, \mn@doi [\apjl] {10.3847/2041-8213/ab9742}, \href
  {https://ui.adsabs.harvard.edu/abs/2020ApJ...896L..30E} {896, L30}

\bibitem[\protect\citeauthoryear{{Hewish}, {Bell}, {Pilkington}, {Scott}  \&
  {Collins}}{{Hewish} et~al.}{1968}]{hewish1968}
{Hewish} A.,  {Bell} S.~J.,  {Pilkington} J.~D.~H.,  {Scott} P.~F.,   {Collins}
  R.~A.,  1968, \mn@doi [\nat] {10.1038/217709a0}, \href
  {https://ui.adsabs.harvard.edu/abs/1968Natur.217..709H} {217, 709}

\bibitem[\protect\citeauthoryear{{Hobbs}, {Edwards}  \& {Manchester}}{{Hobbs}
  et~al.}{2006}]{hobbs2006}
{Hobbs} G.~B.,  {Edwards} R.~T.,   {Manchester} R.~N.,  2006, \mn@doi [\mnras]
  {10.1111/j.1365-2966.2006.10302.x}, \href
  {https://ui.adsabs.harvard.edu/abs/2006MNRAS.369..655H} {369, 655}

\bibitem[\protect\citeauthoryear{{Hotan}, {van Straten}  \&
  {Manchester}}{{Hotan} et~al.}{2004}]{hotan2004}
{Hotan} A.~W.,  {van Straten} W.,   {Manchester} R.~N.,  2004, \mn@doi [\pasa]
  {10.1071/AS04022}, \href
  {https://ui.adsabs.harvard.edu/abs/2004PASA...21..302H} {21, 302}

\bibitem[\protect\citeauthoryear{{Hu} et~al.,}{{Hu} et~al.}{2020}]{hu2020}
{Hu} C.-P.,  et~al., 2020, \mn@doi [\apj] {10.3847/1538-4357/abb3c9}, \href
  {https://ui.adsabs.harvard.edu/abs/2020ApJ...902....1H} {902, 1}

\bibitem[\protect\citeauthoryear{{Karuppusamy} et~al.,}{{Karuppusamy}
  et~al.}{2020}]{karu2020}
{Karuppusamy} R.,  et~al., 2020, The Astronomer's Telegram, \href
  {https://ui.adsabs.harvard.edu/abs/2020ATel13553....1K} {13553, 1}

\bibitem[\protect\citeauthoryear{{Kaspi} \& {Beloborodov}}{{Kaspi} \&
  {Beloborodov}}{2017}]{kaspi2017}
{Kaspi} V.~M.,  {Beloborodov} A.~M.,  2017, \mn@doi [\araa]
  {10.1146/annurev-astro-081915-023329}, \href
  {https://ui.adsabs.harvard.edu/abs/2017ARA&A..55..261K} {55, 261}

\bibitem[\protect\citeauthoryear{Keogh \& Ratanamahatana}{Keogh \&
  Ratanamahatana}{2005}]{keogh2005}
Keogh E.,  Ratanamahatana C.~A.,  2005, \mn@doi [Knowledge and Information
  Systems] {10.1007/s10115-004-0154-9}, 7, 358

\bibitem[\protect\citeauthoryear{{Kouveliotou} et~al.,}{{Kouveliotou}
  et~al.}{1987}]{kou87}
{Kouveliotou} C.,  et~al., 1987, \mn@doi [\apjl] {10.1086/185029}, \href
  {https://ui.adsabs.harvard.edu/abs/1987ApJ...322L..21K} {322, L21}

\bibitem[\protect\citeauthoryear{{Kramer}, {Stappers}, {Jessner}, {Lyne}  \&
  {Jordan}}{{Kramer} et~al.}{2007}]{kramer2007}
{Kramer} M.,  {Stappers} B.~W.,  {Jessner} A.,  {Lyne} A.~G.,   {Jordan} C.~A.,
   2007, \mn@doi [\mnras] {10.1111/j.1365-2966.2007.11622.x}, \href
  {https://ui.adsabs.harvard.edu/abs/2007MNRAS.377..107K} {377, 107}

\bibitem[\protect\citeauthoryear{{Levin} et~al.,}{{Levin}
  et~al.}{2019}]{levin2018}
{Levin} L.,  et~al., 2019, \mn@doi [\mnras] {10.1093/mnras/stz2074}, \href
  {https://ui.adsabs.harvard.edu/abs/2019MNRAS.488.5251L} {488, 5251}

\bibitem[\protect\citeauthoryear{{Li} et~al.,}{{Li} et~al.}{2021}]{li2021}
{Li} C.~K.,  et~al., 2021, Nature Astronomy, \href
  {https://ui.adsabs.harvard.edu/abs/2021NatAs...5..378L} {5, 378}

\bibitem[\protect\citeauthoryear{{Lower}, {Shannon}, {Johnston}  \&
  {Bailes}}{{Lower} et~al.}{2020}]{Lower2020}
{Lower} M.~E.,  {Shannon} R.~M.,  {Johnston} S.,   {Bailes} M.,  2020, \mn@doi
  [\apjl] {10.3847/2041-8213/ab9898}, \href
  {https://ui.adsabs.harvard.edu/abs/2020ApJ...896L..37L} {896, L37}

\bibitem[\protect\citeauthoryear{{Lu}, {Kumar}  \& {Zhang}}{{Lu}
  et~al.}{2020}]{lu2020}
{Lu} W.,  {Kumar} P.,   {Zhang} B.,  2020, \mn@doi [\mnras]
  {10.1093/mnras/staa2450}, \href
  {https://ui.adsabs.harvard.edu/abs/2020MNRAS.498.1397L} {498, 1397}

\bibitem[\protect\citeauthoryear{{Lyne}, {Hobbs}, {Kramer}, {Stairs}  \&
  {Stappers}}{{Lyne} et~al.}{2010}]{lyne2010}
{Lyne} A.,  {Hobbs} G.,  {Kramer} M.,  {Stairs} I.,   {Stappers} B.,  2010,
  \mn@doi [Science] {10.1126/science.1186683}, \href
  {https://ui.adsabs.harvard.edu/abs/2010Sci...329..408L} {329, 408}

\bibitem[\protect\citeauthoryear{{Manchester} et~al.,}{{Manchester}
  et~al.}{2013}]{manchester2013}
{Manchester} R.~N.,  et~al., 2013, \mn@doi [\pasa] {10.1017/pasa.2012.017},
  \href {https://ui.adsabs.harvard.edu/abs/2013PASA...30...17M} {30, e017}

\bibitem[\protect\citeauthoryear{Morozova, Ahmedov  \& Zanotti}{Morozova
  et~al.}{2011}]{morozova2011}
Morozova V.,  Ahmedov B.,   Zanotti O.,  2011, \mn@doi [Proceedings of the
  International Astronomical Union] {10.1017/s1743921312013348}, 7, 359

\bibitem[\protect\citeauthoryear{{Olausen} \& {Kaspi}}{{Olausen} \&
  {Kaspi}}{2014}]{olausen2014}
{Olausen} S.~A.,  {Kaspi} V.~M.,  2014, \mn@doi [\apjs]
  {10.1088/0067-0049/212/1/6}, \href
  {https://ui.adsabs.harvard.edu/abs/2014ApJS..212....6O} {212, 6}

\bibitem[\protect\citeauthoryear{{Palmer} et~al.,}{{Palmer}
  et~al.}{2005}]{palmer2005}
{Palmer} D.~M.,  et~al., 2005, \mn@doi [\nat] {10.1038/nature03525}, \href
  {https://ui.adsabs.harvard.edu/abs/2005Natur.434.1107P} {434, 1107}

\bibitem[\protect\citeauthoryear{Pedregosa et~al.,}{Pedregosa
  et~al.}{2011}]{scikit-learn}
Pedregosa F.,  et~al., 2011, Journal of Machine Learning Research, 12, 2825

\bibitem[\protect\citeauthoryear{{Perera}, {Stappers}, {Weltevrede}, {Lyne}  \&
  {Bassa}}{{Perera} et~al.}{2015}]{perera2015}
{Perera} B.~B.~P.,  {Stappers} B.~W.,  {Weltevrede} P.,  {Lyne} A.~G.,
  {Bassa} C.~G.,  2015, \mn@doi [\mnras] {10.1093/mnras/stu2187}, \href
  {https://ui.adsabs.harvard.edu/abs/2015MNRAS.446.1380P} {446, 1380}

\bibitem[\protect\citeauthoryear{{Perera}, {Stappers}, {Weltevrede}, {Lyne}  \&
  {Rankin}}{{Perera} et~al.}{2016}]{perera2016}
{Perera} B.~B.~P.,  {Stappers} B.~W.,  {Weltevrede} P.,  {Lyne} A.~G.,
  {Rankin} J.~M.,  2016, \mn@doi [\mnras] {10.1093/mnras/stv2403}, \href
  {https://ui.adsabs.harvard.edu/abs/2016MNRAS.455.1071P} {455, 1071}

\bibitem[\protect\citeauthoryear{{Petroff}, {Hessels}  \& {Lorimer}}{{Petroff}
  et~al.}{2021}]{petroff21}
{Petroff} E.,  {Hessels} J.~W.~T.,   {Lorimer} D.~R.,  2021, arXiv e-prints,
  \href {https://ui.adsabs.harvard.edu/abs/2021arXiv210710113P} {p.
  arXiv:2107.10113}

\bibitem[\protect\citeauthoryear{{Rajwade} et~al.,}{{Rajwade}
  et~al.}{2020}]{rajwade2020atel}
{Rajwade} K.,  et~al., 2020, The Astronomer's Telegram, \href
  {https://ui.adsabs.harvard.edu/abs/2020ATel13554....1R} {13554, 1}

\bibitem[\protect\citeauthoryear{{Rajwade}, {Perera}, {Stappers}, {Roy},
  {Karastergiou}  \& {Rankin}}{{Rajwade} et~al.}{2021}]{rajwade2021}
{Rajwade} K.~M.,  {Perera} B.~B.~P.,  {Stappers} B.~W.,  {Roy} J.,
  {Karastergiou} A.,   {Rankin} J.~M.,  2021, \mn@doi [\mnras]
  {10.1093/mnras/stab1942}, \href
  {https://ui.adsabs.harvard.edu/abs/2021MNRAS.506.5836R} {506, 5836}

\bibitem[\protect\citeauthoryear{{Rankin}, {Rodriguez}  \& {Wright}}{{Rankin}
  et~al.}{2006}]{rankin2006}
{Rankin} J.~M.,  {Rodriguez} C.,   {Wright} G. A.~E.,  2006, \mn@doi [\mnras]
  {10.1111/j.1365-2966.2006.10512.x}, \href
  {https://ui.adsabs.harvard.edu/abs/2006MNRAS.370..673R} {370, 673}

\bibitem[\protect\citeauthoryear{{Rea}, {Pons}, {Torres}  \& {Turolla}}{{Rea}
  et~al.}{2012}]{rea2012}
{Rea} N.,  {Pons} J.~A.,  {Torres} D.~F.,   {Turolla} R.,  2012, \mn@doi
  [\apjl] {10.1088/2041-8205/748/1/L12}, \href
  {https://ui.adsabs.harvard.edu/abs/2012ApJ...748L..12R} {748, L12}

\bibitem[\protect\citeauthoryear{{Rickett}}{{Rickett}}{1977}]{rickett1977}
{Rickett} B.~J.,  1977, \mn@doi [\araa] {10.1146/annurev.aa.15.090177.002403},
  \href {https://ui.adsabs.harvard.edu/abs/1977ARA&A..15..479R} {15, 479}

\bibitem[\protect\citeauthoryear{{Stamatikos} et~al.,}{{Stamatikos}
  et~al.}{2020}]{stamatikos2020}
{Stamatikos} M.,  et~al., 2020, GRB Coordinates Network, \href
  {https://ui.adsabs.harvard.edu/abs/2020GCN.27384....1S} {27384, 1}

\bibitem[\protect\citeauthoryear{{Torne} et~al.,}{{Torne}
  et~al.}{2020}]{torne2020}
{Torne} P.,  et~al., 2020, The Astronomer's Telegram, \href
  {https://ui.adsabs.harvard.edu/abs/2020ATel14001....1T} {14001, 1}

\bibitem[\protect\citeauthoryear{{Vinyaikin}}{{Vinyaikin}}{2007}]{vinyaikin2007}
{Vinyaikin} E.~N.,  2007, \mn@doi [Astronomy Reports]
  {10.1134/S1063772907070062}, \href
  {https://ui.adsabs.harvard.edu/abs/2007ARep...51..570V} {51, 570}

\bibitem[\protect\citeauthoryear{{Wilms}, {Allen}  \& {McCray}}{{Wilms}
  et~al.}{2000}]{Wilms}
{Wilms} J.,  {Allen} A.,   {McCray} R.,  2000, \mn@doi [\apj] {10.1086/317016},
  \href {https://ui.adsabs.harvard.edu/abs/2000ApJ...542..914W} {542, 914}

\makeatother
\end{thebibliography}




\appendix
\section{List of observing dates with the Effelsberg, Lovell and MkII radio telescopes.}

\begin{table}
\caption{Observing dates of Lovell, Effelsberg and MKII telescopes for the observing campaign of Swift~J1818.0-1607. Full table can be found in the supplementary materials.}
\label{tab:obs_params}
\centering
\begin{tabular}{l c c c }
\hline\hline
Telescope  &  Topocentric MJDs & Frequency & Bandwidth \\ [0.5ex] 
        &  & (MHz) & (MHz)\\ [0.5ex]
\hline
Effelsberg  & 59117.8092855632 & 6000 & 4000 \\  
            & 59159.6070832431 & 6000 & 4000 \\
            & 59239.3812274257 & 6000 & 4000 \\
            & 59400.9854992102 & 6000 & 4000\\
 
\hline

MKII    & 59053.0416885662 & 1532 & 336 \\
        & 59053.8080755006 & 1532 & 336 \\
        & 59054.0374571576 & 1532 & 336 \\
        & 59054.8047174147 & 1532 & 336 \\
        & 59055.0341480019 & 1532 & 336 \\
        & 59055.8020458138 & 1532 & 336 \\
        & 59056.0331832900 & 1532 & 336 \\
\hline
    Lovell  & 59061.0192579544  & 1532 & 336\\
        & 59066.0168184365 & 1532 & 336\\
        & 59073.0144275636 & 1532 & 336 \\
        & 59103.8710138623 & 1532 & 336 \\
        & 59104.882680093 & 1532 & 336 \\
        & 59105.7356694333 & 1532 & 336 \\
        & 59105.8973174436 & 1532 & 336 \\
 \hline
\end{tabular}
\end{table}


\bsp	
\label{lastpage}
\end{document}